\documentclass[prd,preprint,tightenlines,floatfix,showpacs,preprintnumbers,nofootinbib,eqsecnum]{revtex4}
 \usepackage[dvips,final]{graphicx}
  \usepackage{amssymb}
   \usepackage{amsmath}
    \usepackage{amsfonts}
     \usepackage{epsfig}
      \usepackage{bm}

\begin{document}

\title{Dark Matter in the Standard Model Extension\\ with Singlet Quark}

\author{Vitaly Beylin}
\affiliation{Research Institute of Physics, Southern Federal University,
344090 Rostov-on-Don, Pr. Stachky 194, Russian Federation\vspace{1cm}\footnote{vitbeylin@gmail.com}}

\author{Vladimir Kuksa}
\affiliation{Research Institute of Physics, Southern Federal
University, 344090 Rostov-on-Don, Pr. Stachky 194, Russian
Federation\vspace{1cm}\footnote{vkuksa47@mail.ru}}

\begin{abstract}
We analyze the possibility of hadron Dark Matter carriers consisting of singlet quark and the light standard one. It is shown that stable singlet quarks generate effects of new physics
which do not contradict to restrictions from precision electroweak  data. The neutral and charged pseudoscalar low-lying states are interpreted as the Dark Matter particle and its
mass-degenerated partner. We evaluated their masses and lifetime of the charged component, and describe the potential asymptotes of low-energy interactions of these particles with
nucleons and with each other. Some peculiarities of Sommerfeld enhancement effect in the annihilation process are also discussed.
\end{abstract}

 \maketitle

\section{Introduction}
The problem of Dark Matter (DM) explanation has been in the center of fundamental physics attention for a long time. The existence of the DM is followed from astrophysical data and
remains the essential phenomenological evidences of New Physics' manifestations beyond the Standard Model (SM) \cite{1,2}. An appropriate candidates as DM carriers should be stable
particles which weakly interact with ordinary matter (so called, WIMPs). Such particles usually are considered in the framework of supersymmetric, hypercolor  or other extensions of
the SM (see, for instance, review \cite{3}). The last experimental rigid restrictions on cross section of spin-independent WIMP-nucleon interaction \cite{4} exclude many variants of WIMPs
 as the DM carriers. So, another scenarios are discussed in literature, such as quarks from fourth generation, hyper-colour quarks, dark atoms, axions and so on \cite{3}. In spite of some
 theoretical peculiarities, the possibility of hadronic DM is not excluded and considered, for example, in Refs.~\cite{5}-\cite{10}. The possibility of new hadrons existence, which can be
 interpreted as carriers of the DM, was analyzed in detail within the framework of the SM chiral-symmetric extension \cite{10}.

Principal feature of the hadronic DM structure is that the strong interaction of new stable quarks with standard ones leads to the formation of neutral stable meson or baryon heavy
states. Such scenario can be realized in the extensions of the SM with extra generation \cite{5}-\cite{9}, in mirror and chiral-symmetric models \cite{10,16} or in extensions with
singlet quark \cite{11}-\cite{15}. The second variant was detally considered in Ref.~\cite{10}, where the quark structure and low-energy phenomenology of new heavy hadrons were
described. It was shown that the scenario does not contradict to cosmochemical data, cosmological tests and known restrictions for new physics effects. However, the explicit
realization of the chiral-symmetric scenario faces with some theoretical troubles, which can be eliminated with the help of artificial assumptions. The extensions of SM with fourth
generation and their phenomenology were considered during last decades in spite of strong experimental restrictions which, for instance, follows from invizible Z-decay channel, unitary
condition for CM-matrix, FCNC etc. The main problem of 4th generation is the contribution of new heavy quarks to the Higgs boson decays \cite{10.9}. The contribution of new heavy
quarks to vector boson coupling may be compensated by the contribution of 50 GeV neutrino \cite{13a,Ilyin,Novikov}, however, such assumption looks as artificial. In this paper, we
analyze the hypothesis of hadronic Dark Matter which follows from the SM extension with singlet quark.

The paper is organised as follows. In the second section we describe the extension of the SM with singlet quark and consider the restrictions on its phenomenology, following from
precision electroweak data. Quark composition and interaction of new hadrons with the standard ones at low energies is analyzed in the third section. The masses of new hadrons,
decay properties of charged partner of the DM carrier and annihilation cross section are analyzed in the fourth section.

\section{Standard Model Extension with Stable Singlet quark}

There is a wide class of high-energy extensions of the SM with singlet quarks which are discussed during many decades. Here, we consider the simplest extension of the SM with singlet
quarks as the framework for description of the DM carrier. Singlet (or vector-like) quark is defined as fermion with standard $U_Y(1)$ and $SU_C(3)$ gauge interactions but it is
singlet under $SU_W(2)$ transformations. The low-energy phenomenology of both down- and up-type quarks (D and U) was considered in detail in large number of works (see, for instance,
\cite{9a}, \cite{Eberhardt}, \cite{Botella}, \cite{Kumar} and references therein). As a rule, singlet quark is supposed as unstable due to the mixing with the ordinary ones. This
mixing leads to the FCNC appearing at the tree level. As a consequences, we get an additional contributions into rare processes, such as rare lepton and semy-lepton decays, and mixing
in the systems of neutral mesons ($M^0-\bar{M}^0$ oscillations). The current experimental data on New Fhysics phenomena give rigid restrictions for the angles of ordinary-singlet quark
mixing. In this work, we consider alternative aspect of the extensions with singlet quark $Q$, namely, the scenario with the absence of such mixing.
 As a result, we get stable singlet quark which have no the decay channels due to absence of non-diagonal $Q$-quark currents. More exactly, due to confinement the singlet
 quark forms bound states with the ordinary ones, for instance $(Qq)$, and the lightest state is stable. In this work, we consider some properties of such particles and
 analyze the possibility to interpret the stable neutral meson $M^0=(\bar{Q}q)$ as the DM carrier.

Now, we examine the minimal variants of the SM extension with singlet quark $Q_A$, where subscript $A=U,D$ denotes up- or down- type with charge $q=2/3,-1/3$.
According to the definition, the
field $Q$ is singlet with respect to $SU_W(2)$ group and has standard transformations under abelian $U_Y(1)$ and color $SU_C(3)$ groups. So, the minimal additional gauge-invariant
Lagrangian has the form:
\begin{equation}\label{2.1}
L_Q=i\bar{Q}\gamma^{\mu}(\partial_{\mu}-ig_1 \frac{Y}{2} V_{\mu} -ig_s \frac{\lambda_a}{2}G^a_{\mu})Q - M_Q \bar{Q} Q,
\end{equation}
where $Y/2=q$ is charge in the case of singlet $Q$, and $M_Q$ denotes phenomenological mass of quark. Note, singlet quark (SQ) can not get mass term from the standard Higgs mechanism
because the Higgs doublet is fundamental representation of $SU(2)$ group. Abelian part of the interaction Lagrangian (\ref{2.1}), which will be used in further considerations,
includes the interactions with physical photon $A$ and $Z$ boson:
\begin{equation}\label{2.2}
L_Q^{int}=g_1 q V_{\mu}\bar{Q}\gamma^{\mu} Q = q g_1 (c_w A_{\mu}- s_w Z_{\mu})\bar{Q}\gamma^{\mu} Q,
\end{equation}
where $c_w=\cos\theta_w$, $s_w=\sin\theta_w$, $g_1 c_w=e$ and $\theta_w$ is Weinberg angle of mixing. Note, the left and right parts of the singlet field $Q$ have the same
transformation properties, interaction (\ref{2.2}) has vector-like (chiral-symmetric) form and singlet quark usually is named vector-like quark \cite{Botella,Kumar}.

First of all, we should take into account direct and indirect restrictions on New Physics (NF) manifestations which follow from the precision experimental data. The additional chiral
quarks, for instance from standard fourth generation, are excluded at the $5\,\sigma$ level by LHC data on Higgs searches \cite{Eberhardt}. As the  vector-like (non-chiral) singlet
fermions do not receive their masses from a Higgs doublet, they are allowed by existing experimental data on Higgs physics. The last limits on new colored fermions follow from the jets
data from the LHC \cite{Llorente}. The corresponding limits for effective colored factors $n_{eff}=2,3,6$ are about 200 GeV, 300 GeV, 400 GeV. Note, these limits are much less then the
estimation of quark mass which follows from the DM analysis (see the fourth section). Theoretical and experimental situation for long-lived heavy quarks was considerably discussed in
the review \cite{9a}, where it was noted that vector-like new heavy quarks can elude experimental contraints from LHC.

Indirect limits on new fermions follow from precision electroweak measurements of the effects, such as flavor-changing neutral currents (FCNC) and vector boson polarizations, which
take place at the loop level in the SM. Because we consider the case of stable singlet quark, there are no mixing with ordinary quarks and, consequently, FCNC effects are absent. The
NF manifestations in polarization effects of gauge bosons $\gamma,\,Z,\,W$ are usually described by oblique Peskin-Takeuchi parameters \cite{Peskin} (PT parameters). From
Eq.~(\ref{2.2}), it follows that the singlet quark gives non-zero contributions into polarizations of $\gamma$ and $Z$-bosons which are described by the values of
$\Pi_{\gamma\gamma},\,\Pi_{\gamma Z},\,\Pi_{ZZ}$. As $W$-boson does not interact with the SQ, corresponding contribution into polarizaton operator is zero, $\Pi_{WW}=0$. These
parameters are expressed in terms of vector bosons polarizations $\Pi_{ab}(p^2)$, where $a,b=W, Z, \gamma$. Here, we use the definition $\Pi_{\mu\nu}(p^2)=p_{\mu}p_{\nu}P(p^2) +
g_{\mu\nu}\Pi(p^2)$ and the expressions for PT oblique parameters  from \cite{Burgess}. In the case under consideration, $\Pi_{ab}(0)=0$ and PT parameters can be represented by the
following expressions:
\begin{align}\label{2.3}
\alpha S=&4s^2_w c^2_w[\frac{\Pi_{ZZ}(M^2_Z,M^2_U)}{M^2_Z}-\frac{c^2_w-s^2_w}{s_w c_w}\Pi^{'}_{\gamma Z}(0,m^2_U)-\Pi^{'}_{\gamma\gamma}(0,M^2_U)];\notag\\
\alpha U=&-4s^2_w[c^2_w\frac{\Pi_{ZZ}(M^2_Z,M^2_U)}{M^2_Z}+2s_w c_w \Pi^{'}_{\gamma Z}(0,M^2_U)+s^2_w\Pi^{'}_{\gamma\gamma}(0,M^2_U)];\notag\\
\alpha T=&-\frac{\Pi_{ZZ}(0,M^2_U)}{M^2_Z}=0;\,\,\, \alpha V=\Pi^{'}_{ZZ}(M^2_Z,M^2_U)-\frac{\Pi_{ZZ}(M^2_Z,M^2_U)}{M^2_Z};\notag\\
\alpha W=&0\,\,(W\sim \Pi_{WW}=0);\,\,\,\alpha X=-s_w c_w [\frac{\Pi_{\gamma Z}(M^2_Z,M^2_U)}{M^2_Z}-\Pi^{'}_{\gamma Z}(0,M^2_U)].
\end{align}
In (\ref{2.3}) polarizations $\Pi_{ab}(p^2,M^2_U)$, where $a,b=\gamma,Z$, in one-loop approach can be represented in simple form (for the case of SQ with $q=2/3$):
\begin{align}\label{2.4}
\Pi_{ab}(p^2,M^2_U)=&\frac{g_1^2}{9\pi^2}k_{ab}F(p^2,M^2_U);\,\,\,k_{ZZ}=s^2_w,\,k_{\gamma\gamma}=c^2_w,\,k_{\gamma Z}=-s_w c_w;\notag\\
F(p^2,M^2_U)=&-\frac{1}{3}p^2+2M^2_U+2A_0(M^2_U)+(p^2+2M^2_U)B_0(p^2,M^2_U).
\end{align}
In Eqs.~(\ref{2.4}) the function $F(p^2,M^2_U)$ contains divergent terms in the one-point, $A_0(M^2_U)$, and two-point, $B_0(p^2,M^2_U)$, Veltman functions which are exactly
compensated in physical parameters (\ref{2.3}). Using standard definitions of the functions $A_0(M^2_U)$ and $B_0(p^2,M^2_U)$ and the equality  $B^{'}_0(0,M^2_U)=M^2_U/6$, by
straightforward calculations we get simple expressions for oblique parameters:
\begin{equation}\label{2.5}
S=-U=\frac{16s^4_w}{9\pi}[-\frac{1}{3}+2(1+2\frac{M^2_Q}{M^2_Z})(1-\sqrt{\beta}\arctan\frac{1}{\sqrt{\beta}})],
\end{equation}
where $\beta=4M^2_Q/M^2_Z -1$. We check that in the limit $M^2_Q/M^2_Z\to \infty$ the values of $S$ and $U$ go to zero as $\sim M^2_Z/M^2_Q$ in accordance with well-known results for
the case of vector-like interactions \cite{2,Burgess}. From Eq.~(\ref{2.5}) it follows that beginning from $M_Q=500$ GeV the parameter $S<10^{-2}$ and the rest non-zero parameters have
near the same values.
 These values significantly less the current experimental limits \cite{PDG}:
$S=0.00 +0.11(-0.10),\,\,\,U=0.08\pm 0.11,\,\,\,T=0.02+0.11(-0.12)$, that is the scenario with up-type singlet quark satisfy to the restrictions on indirect manifestations of heavy new
fermions. Note, parameters $V,\,W,\,X$ describe the contributions of new fermions with masses close to the electro-weak scale. In the case of down-type singlet quark, having charge
$q=-1/3$, the contributions into all polarizations and, consequently, into PT parameters are four times smaller.

In the quark-gluon phase (QGP) of the Universe evolution, stable SQ interacts with standard quarks through exchanges by gluons $g$, $\gamma$ and $Z$ according to Eq.~(\ref{2.1}). So,
we have large cross-section for annihilation into gluons and quarks, $Q\bar{Q} \to gg$ and $Q\bar{Q}\to q\bar{q}$ correspondingly, and also small additional contributions in
electroweak channels $Q\bar{Q} \to \gamma\gamma,\,ZZ$. These cross sections can be simply derived from the known expressions for the processes $gg\to Q\bar{Q}$ and $q\bar{q}\to
Q\bar{Q}$ (see review in Ref.~\cite{PDG}) by time inversion.
 Two-gluon cross section in the low-energy limit looks like:
\begin{equation}\label{2.6}
 \sigma(U\bar{U}\to gg)=\frac{14\pi}{3}\frac{\alpha^2_s}{v_r M^2_U},
\end{equation}
where $M_U$ is mass of $U$-quark and $\alpha_s=\alpha_s(M_U)$ is strong coupling at the corresponding scale.  Two-quark channel in the massless limit $m_q\to 0$ is as follows:
\begin{equation}\label{2.7}
\sigma(U\bar{U}\to q\bar{q})=\frac{2\pi}{9}\frac{\alpha^2_s}{v_r M^2_U}.
\end{equation}
So, the two-gluon channel dominates. We should note, that the cross section of SQ -annihilation is suppressed by large $M_U$ in comparison with the annihilation of standard quarks.

After the transition from quark-gluon plasma to hadronization stage, the singlet quarks having standard strong interactions (gluon exchange), form coupled states with ordinary quarks.
New heavy hadrons can be constructed as coupled states which consist of heavy stable quark $Q$ and a light quark from the SM quark sector. Here, we consider the simplest two-quark
states, neutral and charged mesons. The lightest of them, for instance neutral meson $M=(\bar{Q}q)$, is stable and can be considered as the carrier of cold Dark Matter. Possibility of
existence of heavy stable hadrons was carefully analyzed in \cite{10}, where it was shown that this hypothesis does not contradict to cosmochemical data and cosmological test. This
conclusion was based on the important property of new hadron, namely, repulsive strong interaction with nucleons at large distances. The effect will be qualitatively analyzed for the
case of $MM$ and $MN$ interactions in the next section.

\section{Quark composition of new hadrons\\ and their interactions with nucleons}

At the hadronization stage, heavy SQ form the coupled states with the ordinary light quarks. Classification of these new heavy hadrons was considered in Ref.~\cite{10}, where quark
composition of two-quark (meson) and three-quark (fermion) states was represented for the case of up- and down-types of quark $Q$. Stable and long-lived new hadrons are divided into
three families of particles with characteristic values of masses M, 2M and 3M, where M is the mass of $Q$-quark. Quantum numbers and quark content of these particles for the case of
up-type quark $Q=U$ are represented in Table 1.
\begin{center}
{Table 1. Characteristics of $U$-type hadrons}
\end{center}
\begin{center}
\begin{tabular}{||l|l|l|l||}
\hline $J^P=0^-$         &$T=\frac{1}{2}$    &$M=(M^0\,M^-)$   &$M^0=\bar{U}u$,\, $M^-=\bar{U}d$\\ \hline
$J=\frac{1}{2}$   &$T=1$              &$B_1=(B_1^{++}\,B_1^+\,B_1^0)$ &$B_1^{++}=Uuu,B_1^+=Uud,B_1^0=Udd$\\
\hline $J=\frac{1}{2}$    &$T=\frac{1}{2}$     &$B_2=(B^{++}_2\,B^+_2)$ &$B^{++}_2=UUu,B^+_2=UUd$\\ \hline $J=\frac{3}{2}$ &$T=0$ &$(B^{++}_3)$ &$B^{++}_3=UUU$\\ \hline
\end{tabular}
\end{center}
Some states in Table 1 were also considered in Ref.~\cite{9a} for the case of long-lived vector-like heavy quark and in Ref.~\cite{10.8}, where $U$-type quark belong to the sequential
4-th generation. In Ref.~\cite{13d}, there were considered an important property of suppression of hadronic interaction of heavy quark systems containing three new quarks, like $(UUU)$
states. This model has $SU(3)\times SU(2)\times SU(2)\times U(1)$ symmetry and offers a novel alternative for the DM carriers --- they can be an electromagnetically bound states made
of terafermions. The charged $M^-$ and neutral $M^0$ particles can manifest themselves in cosmic rays and as carrier of the DM. In Refs.~\cite{7,8,9} a possibility is discussed that
new stable charged hadrons exist but are hidden from detection, being bounded inside neutral dark atoms. For instance, stable particles with charge $Q=-2$ can be bound with primordial
helium.

Interactions of the baryon-type particles $B_1$ and $B_2$ (the second and third line in Table 1) are similar to the nucleonic ones, and they may compose atomic nuclei together with
nucleons. As it was demonstrated in Ref.~\cite{10}, this circumstance does not prevent the $B_1$ and $B_2$ burn out in the course of cosmochemical evolution. There are no problems also
with interaction of $B_3$ isosinglet  with nucleons which proceeds mainly through exchange by mesons, $\eta$ and $\eta^{'}$. The constants of such interactions, as it follows from the
quark model of the mesonic exchange (see Ref.~\cite{10}), is not a large one, i.e. $B_3N$ interaction is suppressed in comparison with the $NN$ interaction.

There is another type of hypothetical hadrons which possess analogous properties of strong interactions. They are constructed from stable quark of the down-type (D-quark) with $Q =
-1/3$ electric charge. Quantum numbers and quark content of these particles are represented in Table 2 (see the corresponding analysis and comments in Ref.~\cite{10}).
\begin{center}
{Table 2. Characteristics of $D$-type hadrons.}
\end{center}
\begin{center}
\begin{tabular}{||l|l|l|l||}
\hline $J^P=0^-$         &$T=\frac{1}{2}$    &$M_D=(M^+_D\,M^0_D)$   &$M^+_D=\bar{D}u$,\, $M^0_D=\bar{D}d$\\ \hline
$J=\frac{1}{2}$   &$T=1$              &$B_{1D}=(B_{1D}^+\,B_{1D}^0\,B_{1D}^-)$ &$B_{1D}^{+-}=Duu(Ddd),B_{1D}^0=Dud$\\
\hline $J=\frac{1}{2}$    &$T=\frac{1}{2}$     &$B_{2D}=(B^0_{2D}\,B^-_{2D})$ &$B^0_{2D}=DDu,B^-_{2D}=DDd$\\ \hline $J=\frac{3}{2}$ &$T=0$ &$(B^-_{3D})$ &$B^-_{3D}=DDD$\\ \hline
\end{tabular}
\end{center}
In this table, the states $M^+_D,\,B^0_{1D},\,B^0_{2D},\,B^-_{3D}$ are stable. Particles possessing a similar quark composition appear in various high-energy generalizations of SM, in
which $D$-quark is a singlet with respect to weak interactions group. For example, each quark-lepton generation in $E(6)\times E(6)$ -model contains two singlet $D$-type quarks. This
quark appears, also, from the Higgs sector in supersymmetric generalization of $SU(5)$ Great Unification model.  As a rule, with a reference to cosmological restrictions it is assumed
that new hadrons are unstable due to the mixing of singlet $D$-quarks with the standard quarks of the down type. Note, the consequences for cosmochemical evolution, caused by existence
of the hypothetical stable
 $U$- and $D$-types hadrons, are very different.

Cosmochemical evolution of new hadrons at hadronization stage was qualitatively studied both for $U$ and $D$ cases in \cite{10}. A very important conclusion was arrived from this
analysis - baryon asymmetry in new quark sector must exist and has a sign opposite to asymmetry in standard quark sector (quarks $U$ disappear but antiquarks $\bar{U}$ remain). This
conclusion follows from the strong cosmochemical restriction for the ratio ``anomalous/natural'' hydrogen $C\leqslant 10^{-28}$ for $M_Q\lesssim 1\,\mbox{TeV}$ \cite{Smith} and
anomalous helium $C\leqslant 10^{-12} - 10^{-17}$ for $M_Q\leq 10\,\mbox{TeV}$ \cite{Muller}. In our case, the state $B^+_1=(Uud)$ is heavy (anomalous) proton which can form anomalous
hydrogen. At the stage of hadronization, $B^+_1$ can be formed by direct coupling of quarks and as a result of reaction $\bar{M}^0 + N \to B^+_1 + X$, where $X$ is totality of leptons
and photons in the final state. The antyparticles $\bar{B}^+_1$ are burning out due to the reaction $\bar{B}^+_1 + N \to M^0 +X$. The states like $(pM^0)$ can be also manifest itself
as anomalous hydrogen, but as it was shown in \cite{10}, interaction of $p$ and $M^0$ has a  potential barrier at large distances. So, formation of coupled states $(pM^0)$ at low
energies is strongly suppressed. As it follows from the experimental restrictions on anomalous hydrogen and helium \cite{Smith,Muller}, baryon symmetry in extra sector of quarks is not
excluded for the case of super-heavy new quarks with masses $M_Q\gg 1\,\mbox{TeV}$ (see, also, the fourth section). Further, we consider the interaction of new hadrons with nucleons
and their self-interaction in more detail.

At low energies the hadrons interactions can be approximately described by a model of meson exchange in terms of an effective lagrangian. It was shown in \cite{18}, that low-energy
baryon-meson interactions are effectively described by $U(1)\times SU(3)$ gauge theory, where $U(1)$ is the group of semi-strong interaction and $SU(3)$ is group of hadronic unitary
symmetry.
 Effective physical lagrangian which was used for calculation of $MN$ interaction potential is represented in \cite{10}. By straightforward calculations, it was demonstrated there that the
 dominant contribution is resulted from the exchanges by $\rho$ and $\omega$ mesons. This lagrangian at low energies can be applied for analysis both of $MN$ and $MM$ interactions. Here,
 we give the part of lagrangian with vector-meson exchange which will be used for evaluation of the potential:
\begin{align}\label{3.1}
L_{int}&=g_{\omega}\omega^{\mu}\bar{N}\gamma_{\mu}N + g_{\rho}\bar{N}\gamma_{\mu}\hat{\rho}^{\mu}N
       +ig_{\omega M}\omega^{\mu}(M^{\dagger}\partial_{\mu}M-\partial_{\mu}M^{\dagger}M)\notag\\ &+ ig_{\rho M} (M^{\dagger}\hat{\rho}^{\mu}\partial_{\mu}M-
       \partial_{\mu}M^{\dagger}\hat{\rho}^{\mu}M).
\end{align}
In (\ref{3.1}) $N=(p,n),\,M=(M^0,\,M^-),\,M^{\dagger}=(\bar{M}^0,\,M^+)$ and coupling constants are the following \cite{10}:
\begin{align}\label{3.2}
g_{\rho}&=g_{\rho M}=g/2,\,\,\,g_{\omega}=\sqrt{3}g/2\cos\theta ,\,\,\,g_{\omega M}=g/4\sqrt{3}\cos\theta,\notag\\
        &g^2/4\pi\approx 3.16,\,\,\,\cos\theta =0.644.
\end{align}
Note, the one-pion exchange which is dominant in $NN$  interaction is forbidden in the $MM\pi$ -vertex due to parity conservation.

In Born approximation, the potential of interaction and the non-relativistic amplitude of scattering for the case of non-polarized particles are connected by the relation:
\begin{equation}\label{3.3}
U(\vec{r})=-\frac{1}{4\pi^2\mu}\int f(q)\exp(i\,\vec{q}\, \vec{r})\, d^3q,
\end{equation}
where $\mu$ is the reduced mass of scattering particles. For the case of $M$ scattering off nucleons, this potential was calculated in Ref.~\cite{10}, where it was utilized the
relation $f(q)=-2\pi i\mu F(q)$ between nonrelativistic amplitude, $f(q)$, and Feynman amplitude, $F(q)$. As it was shown, contributions of scalar and two-pion exchanges are
suppressed by the factor
$\sim m_N/m_M$. Expressions for potentials of interaction of various pairs from doublets $(M^0,M^-)$ and $(p,n)$ have following form:
\begin{align}\label{3.4}
U(M^0,p;r)&=U(M^-,n;r)\approx U_{\omega}(r)+U_{\rho}(r),\notag\\
U(M^0,n;r)&=U(M^-,p;r)\approx U_{\omega}(r)-U_{\rho}(r).
\end{align}
In Eqs.(\ref{3.4}) the terms $U_{\omega}(r)$ and $U_{\rho}(r)$ are defined by the following expressions:
\begin{equation}\label{3.5}
U_{\omega}=\frac{g^2K_{\omega}}{16\pi\cos^2\theta}\,\frac{1}{r}\,\exp(-\frac{r}{r_{\omega}}),\,\,\,U_{\rho}=\frac{g^2K_{\rho}}{16\pi}\,\frac{1}{r}\,\exp(-\frac{r}{r_{\rho}}),
\end{equation}
where $K_{\omega}=K_{\rho}\approx0.92,\,\,r_{\omega}=1.04/m_{\omega},\,\,r_{\rho}=1.04/m_{\rho}$. Taking into account these values and $m_{\omega}\approx m_{\rho}$, we rewrite
expressions (\ref{3.4}) in a form:
\begin{align}\label{3.6}
U(M^0,p;r)&=U(M^-,n;r)\approx 2.5\,\frac{1}{r}\,\exp(-\frac{r}{r_{\rho}}),\notag\\
U(M^0,n;r)&=U(M^-,p;r)\approx 1.0\,\frac{1}{r}\,\exp(-\frac{r}{r_{\rho}}).
\end{align}
Two consequences can be deduced from the expressions (\ref{3.6}). Firstly, all four pairs of particles have repulsive potential ($U>0$) of interaction at long distances,
where Born approximation is valid. Secondly, due to potential barrier  the DM particles at low energies can not interact with nucleons, i.e. they can not form the coupled
states $(pM^0)$ which manifest itself as anomalous protons. So, they can not be directly detected. To overcome the barrier, nucleons should have energy $\sim 1\, GeV$ or
more and this situation takes place in high energy cosmic rays.

Potential of $MM$ interaction can be also reconstructed with the help of above given method. Here, we determine only the sign of potential which define characteristic (attractive or
repulsive) of interaction at long distances. This characteristic plays crucial role for low-energy collisions of the DM particles and nucleons. To determine the sign of potential we
use the definition of lagrangian in the non-relativistc limit:
\begin{equation}\label{3.7}
L=L_0+L_{int}\,\longrightarrow W_k-U,
\end{equation}
where $W_k$ is kinetic part and $U$ is potential. There is a relation between effective $L_{int}(q)$ and Feynman amplitude $F(q)$: $F(q)=ikL_{int}(q)$, where $k>0$ is real coefficient
depending on the type of particles. As a result, we get equality $signum(U)=signum(iF)$, where amplitude of interaction is determined by one-particle exchange diagrams for the process
$M_1M_2\to M^{'}_1M^{'}_2$. Here, $M=(M^0,\bar{M^0})$ and vertexes are defined by the low-energy lagrangian (\ref{3.1}). With the help of this simple approach, one can check previous
conclusion about repulsive character of $MN$ interactions. First of all it should be noted, that low-energy effective lagrangians of $NM^0$ and $N\bar{M}^0$ have opposite sign due to
different sign of vertexes $\omega M^0 M^0$ and $\omega \bar{M}^0 \bar{M}^0$. This effect can be seen from the differential structure of corresponding part of Lagrangian (\ref{3.1})
and representation of field function of the $M$ -particle in the form:
\begin{align}\label{3.8}
M(x)=&\sum_p \hat{a}^-_p (M)\exp(-ipx)+\hat{a}^+_p(\bar{M})\exp(ipx),\notag\\
M^{\dagger}(x)=&\sum_p \hat{a}^+_p (M)\exp(ipx)+\hat{a}^-_p(\bar{M})\exp(-ipx).
\end{align}
In Eqs.~(\ref{3.8}), $a^{\pm}_p(M)$ and $a^{\pm}_p(\bar{M})$ are the operators of creation and destruction of particles $M$ and antiparticles $\bar{M}$ with momentum $p$. As a result,
we get the vertexes $\omega(q) M^0(p) M^0(p-q)$ and $\omega(q) \bar{M}^0(p) \bar{M}^0(p-q)$ in momentum representation with opposite signs, $L_{int}=\pm g_{\omega M}(2p-q)$,
respectively. This leads to the repulsive and attractive potentials of $N M$ and $N \bar{M}$ low-energy effective interactions via $\omega$ exchange. Thus, the absence of potential
barrier in the last cases give rise to the problem of coupled states $p\bar{M}^0$ (the problem of anomalous hydrogen). As it was noted earlier, to overcome this problem we make the
suggestion that the hadronic DM is baryon asymmetric ($\bar{M}^0$ is absent at low-energy stage of hadronization) or particles $\bar{M}^0$ are superheavy.  Properties of interactions
of baryons $B_1$ and $B_2$ are similar to nucleonic one (the main contribution give one-pion and vector meson exhanges) and together with nucleons they may compose an atomic nuclei.
So, new baryons can form superheavy nuclears which in the process of evolution are concentrated due to gravitation in the center of massive planets or stars.

Further, we have checked that the potential of $M^0M^0$ and $\bar{M}^0\bar{M}^0$ interactions is attractive ($U<0$) for the case of scalar meson exchange and repulsive for the case of
vector meson exchange. Potential of $M^0\bar{M}^0$ scattering has attractive asymptotes both for scalar and vector meson exchanges. Thus, the presence of potential barrier in the
processes of $M^0M^0$ and $\bar{M}^0\bar{M}^0$ scattering depends on the relative contribution of scalar and vector mesons. In the case of $M^0\bar{M}^0$ scattering the total potential
is attractive and this property can lead to increasing of annihilation cross section in an analogy with Sommerfeld effect \cite{SEeff}.

\section{Main properties of new hadrons as the DM carriers}

The mass of heavy quark $M_Q$ and the mass splitting of the charged $M^- $ and neutral $M^0$ mesons, $\delta m = m^- - m^0$, are significant characteristics of these states both for
their physical interpretation and for application in cosmology. In this analysis, we take into consideration standard electromagnetic and strong interactions only. So, some
properties of new mesons doublet $M=(M^0,M^-)$ are analogous to properties of standard mesons consisting of pairs of heavy and light quarks. From experimental data on mass
splitting in neutral-charged meson pairs $K=(K^0,K^{\pm})$, $D=(D^0,D^{\pm})$ and $B=(B^0,B^{\pm})$, it is seen that for down-type mesons $K$ and $B$ the mass-splitting
 $\delta m <0$ while for up-type meson $D$ the value of $\delta m>0$. Such results can be explained by the fitting data on current masses of quarks, $m_d>m_u$, and binding
 energy of the systems $(\bar{Q}u)$ and $(\bar{Q}d)$, where Coulomb contributions have different signs. The absolute value of $\delta m $ for the case of $ K- $ and $ D- $ mesons
 is $O(\mbox{MeV})$, but for $B- $ mesons it is less. Taking into account these data, for the case of up SQ we assume:
\begin{equation}\label{4.1}
\delta m =m(M^-)-m(M^0)>0,\,\,\,\mbox{and}\,\,\, \delta m=O(MeV).
\end{equation}
Then, we conclude that neutral state $M^0=(\bar{U}u)$ is stable and can play the role of the DM carrier. The charged partner $M^-=(\bar{U}d)$ has only one decay channel with very
small phase space:
\begin{equation}\label{4.2}
M^-\to M^0e^-\bar{\nu}_e,\,\,\,(\mbox{if}\,\,\,\delta m>m_e).
\end{equation}
This semileptonic decay is resulted from the weak transition $d\to u+W^-\to u+e^-\bar{\nu}_e$, where heavy quark $\bar{U}$ is considered as spectator. The width of decay can be
calculated in a standard way and final expression for differential width is as follows (see also review by R. Kowalski in \cite{PDG}):
\begin{equation}\label{4.3}
\frac{d\Gamma}{d\omega}=\frac{G^2_F}{48\pi^3}|U_{ud}|^2(m_-+m_0)^2m_0^3(\omega^2-1)^{3/2}G^2(\omega).
\end{equation}
In the case under consideration $m_-\approx m_0$, $\omega =k^0/m_0\approx 1$ and $G(\omega)\approx 1$ (HQS approximation). Here, $G(\omega)$ is equivalent to normalized formfactor
$f_+(q)$, where $q$ is the transferred momentum. In the vector dominance approach this formfactor is defined as $f_+(q)=f_+(0)/(1-q^2/m^2_v)$, where $m_v$ is the mass of vector
intermediate state. So, HQS approximation corresponds to the conditions $q^2\ll m^2_v$ and $f_+(0)\approx 1$ for the case $\omega =k^0/m_0\approx 1$. Using Eq.(\ref{4.3}), for the
total width we get:
\begin{equation}\label{4.4}
\Gamma\approx \frac{G^2_F|U_{ud}|^2 m^5_0}{12\pi^3}\int_1^{\omega_m}(\omega^2-1)^{3/2} d\omega;\,\,\,\omega_m=\frac{m^2_0+m^2_-}{2m_0m_-}.
\end{equation}
After integration, the expression (\ref{4.4}) can be written in the simple form:
\begin{equation}\label{4.5}
\Gamma\approx \frac{G_F^2}{60\pi^3}(\delta m)^5,
\end{equation}
where weak coupling constant is taking at a low-energy scale because of small transferred momentum in the process. From the expression (\ref{4.5}) one can see that the width crucially
depends on the mass splitting, $\Gamma\sim (\delta m)^5$ and does not depend on the mass of meson $M$. For instance, in the interval $\delta m=(1-10)\,\mbox{MeV}$ we get following estimations:
\begin{equation}\label{4.6}
\Gamma\sim (10^{-29}-10^{-24})\,\mbox{GeV};\,\,\,\tau\sim (10^5-10^0)\,\mbox{s}.
\end{equation}
Thus, charged partner of $M^0$, which  is long-lived (metastable), can be directly detected in the processes of $M^0 N-$ collisions with an energetic nucleons, $N$. This conclusion is
in accordance with the experimental evidence of heavy charged metastable particles presence in cosmic rays (see Ref.~\cite{10} and references therein). Note olso, the models of DM with
a long-lived co-annihilation partner are discussed in literature (see, for instance, Refs.~\cite{9a,Khoze}).

Experimental and theoretical premises of new heavy hadron existence were discussed in the Ref.~\cite{10}. With the help of low-energy model of baryon-meson interactions, it was shown
that the potential of $MN$ -interaction has repulsive asymptotics. So, the low-energy particles $M$ do not form coupled states with nucleon and the hypothesis of their existence does
 not contradict to the cosmochemical data.

Now, we estimate the mass of new hadrons which are interpreted as carriers of the DM. The data on Dark Matter relic concentration result to value of the cross section of annihilation
at the level:
\begin{equation}\label{4.7}
(\sigma v_r)^{exp}\approx 10^{-10}\,\,GeV^{-2}.
\end{equation}
Comparing the model annihilation cross section (which depends on the mass) to this value, we estimate the mass of the meson $M^0$. Note, the calculations are fulfilled for the case
of hadron-symmetrical DM, that is, the relic abundance is suggested the same for $M^0$ and $\bar{M}^0$. To escape the contradiction with strong restriction on anomalous helium,
 we should expect the mass of $M^0$ above 10 TeV. Approximate evaluation of the model cross section $\sigma(M^0\bar{M}^0)$ can be fulfilled in spectator approach
 $\sigma(M^0\bar{M}^0)\sim\sigma(U\bar{U})$ considering the light $u$-quarks as spectators. Main contributions to this cross section result from sub-processes
 $U\bar{U}\to gg$ and $U\bar{U}\to q\bar{q}$, where $g$ and $q$ are standard gluon and quark. Corresponding cross sections are represented in the second section
 (Eqs.~(\ref{2.6}) and (\ref{2.7})) and their sum is used for approximate evaluation of the
full annihilation cross section of the processes $M^0\bar{M}^0\to \mbox{hadrons}$. Thus, we can estimate $M_U$ mass from the following approximate equation:
\begin{equation}\label{4.8}
(\sigma v_r)^{exp}\approx \frac{44\pi}{9}\frac{\alpha^2_s}{M^2_U}.
\end{equation}
Now, from (\ref{4.7}) and (\ref{4.8}) we get: $m(M^0)\approx M_U\approx 20\,\mbox{TeV}$ at $\alpha_s=\alpha_s(M_U)$. Note, this value get into the range (10--100) TeV which was
declared for the case of heavy WIMPonium states in Ref.~\cite{Asadi}.

As it was noted in the previous section, attractive potential of $M^0\bar{M}^0$  interaction at long distances can increase the cross section due to the light meson exchange.
This effect leads to Sommerfeld enhancement \cite{SEeff} of the cross section:
\begin{equation}\label{4.9}
\sigma v_r=(\sigma v_r)_0S(\alpha/v),
\end{equation}
where $(\sigma v_r)_0$ is initial cross section which is results from the left side of the expression (\ref{4.8}); $\alpha=g^2/4\pi$ is defined by the effective coupling
according to (\ref{3.2}) and $v=v_r/2$. At $m\ll M\approx M_U$, where $m$ is mass of mesons (the light force carriers), Sommerfeld enhancement (SE) factor can be represented
in the form  \cite{SEeff}:
\begin{equation}\label{4.10}
S(\alpha/v)=\frac{\pi \alpha/v}{1-\exp(-\pi \alpha/v)}\,.
\end{equation}

In our case, the light force carriers are $\omega$- and $\rho$ -mesons and $\alpha\sim 1$ (see (\ref{3.1}) and (\ref{3.2})), so from (\ref{4.10}), we get $10^2 \lesssim S(\alpha/v)/\pi
\lesssim 10^3$ in the interval $10^{-2}>v>10^{-3}$ . In this case, from (\ref{4.8})-(\ref{4.10}) it follows that at $v\sim 10^{-2}$ the mass of new quark $M_U\sim 10^2\,\mbox{TeV}$,
which agree with the evaluation of the mass of baryonic DM in \cite{RanHuo} ($M\sim 100$ TeV). Thus, we get too heavy $M^0$ which can not be detected in the searching for signals of
anomalous hydrogen ($M_{max}\lesssim 1\,\mbox{TeV}$) and anomalous helium ($M_{max}\lesssim 10\,\mbox{TeV}$). Note, however, that in these calculations we take into account the light
mesons only, ($m\ll M_U$), which act at long distances $r\sim m_{\rho}^{-1}$. At short distance, near the radius of coupling state $M^0=(\bar{U}u)$, i.e. at $r\sim M^{-1}_U$, it is
possible the exchange by heavy mesons containing heavy quark $U$, for instance, by vector or scalar $M$ -mesons. In this case, the expression (\ref{4.10}) is not valid because of
$M_{\chi}\sim M_U$, where $M_{\chi}$ is the mass of heavy force carriers. To evaluate SE factor in this case, we use its numerical calculation from \cite{Cirelli}, where iso-contours
of the SE corrections are presented as functions of $y=\alpha M/M_{\chi}$ and $x=\alpha/v$. Then $y\approx 1$, and from Ref.~\cite{Cirelli} (see Fig.1 there) it follows that $S\approx
10$ in the interval $10^{-1}>v>10^{-3}$. As a result, from (\ref{4.8}) and (\ref{4.10}) it follows $M_U\approx 60\,TeV$ which does not change situation crucially. It should be noted,
full description of SE requires an account of weak vector bosons $Z,W$ which interact with light quarks only. Thus, SE effect is formed at various energy regions corresponding to
various distances and has very complicated and vague nature (see, also, Ref.~\cite{Blum}).

\section{Conclusion}

We have analyzed a scenario of the hadronic DM based on the simplest extension of the SM with singlet quark. It was shown in a previous work that the existence of new heavy hadrons
does not contradict to cosmological constraints. Here, we demonstrate that the scenario is in accordance with the precision electroweak restrictions on manifestations of New Physics.
With the help of effective model Lagrangian, we describe the asymptotes of interaction potential at low energies for interactions of new hadrons with nucleons and with each other.
These asymptotics occure both atractive and repulsive for different pairs of interacting particles $N,M$ and their antipaticles. The cosmochemical constrictions on anomalous hydrogen
amd anomalous helium lead to the conclusion that abundance of particles $M$ and antiparticles $\bar{M}$ is strongly asymmetrical, or new hadrons $M$ are superheavy (with mass larger 10
TeV).

Approximate value of the mass-splitting for charged and neutral components was evaluated and lifetime of charged meta-stable hadron component was calculated, it occurs rather large,
$\tau\gg 1$ s. Using the value of the DM relic concentration and the expression for the model cross section of annihilation, mass of the hadronic DM carrier is estimated. The value of
mass without account of SE effect is near 20 TeV and the SE increases it up to an order of $10^2$ TeV. These results agree with the evaluations of mass of baryonic DM, which are
represented in literature (see previous section). So, superheavy new hadrons can not be generated in the LHC experiments and detected in the searching for anomalous hydrogen and
helium. Some peculiarities of Sommerfeld enhancement effect in the process of annihilation are analyzed. It should be underlined, that the model annihilation cross section was
evaluated at the level of sub-processes. So, for the description of the hadronic Dark Matter in more detail it is necessary to clarify the mechanism of annihilation process at various
energy scales.

\section*{Data Availability}
The graphic data used to support the findings of this study are available from the corresponding author upon request.

\section*{Conflict of interest}
The authors declare that they have no conflict of interest.

\section*{Acknowledgments}
The work was supported by Russian Scientific Foundation (RSCF) [Grant No.: 18-12-00213].



\begin{thebibliography}{00}    

\bibitem{1} F. Sannino, ``Conformal Dynamics for TeV Physics and Cosmology'', {\it Acta Physica Polonica B}, vol. 40, no. 12, pp. 3533-3744, 2009.

\bibitem{2} R. Pasechnik, V. Beylin, V. Kuksa, and G. Vereshkov, ``Vector-like technineutron Dark Matter: is a QCD-type Technicolor ruled out by XENON100?'', {\it European Physical Journal C},
vol. 74, no. 2, p. 2728, 2014.

\bibitem{3} M. Khlopov, ``Cosmological Reflection of Particle Symmetry'', {\it Symmetry}, vol. 8, p. 81, 2016.

\bibitem{4} XENON Collab. (E. Aprile et al.), ``First Dark Matter Search Results from the XENON1 Experiment'', {\it Physical Review Letters}, vol. 119, article 181301, 2017.

\bibitem{5} M. Maltoni, V.A. Novikov, L.B. Okun, A.N. Rozanov, and M.I. Vysotsky, {\it Physical Letters B}, vol. 476, no. 1-2, pp. 107-115, 2000.

\bibitem{6} K.M. Belotsky, D. Fargion, M.Yu. Khlopov et al., ``Heavy hadrons of 4th family hidden in our Universe and close to detection'', {\it Gravit. Cosmol. Suppl.}, vol. 11, p. 3, 2005.

\bibitem{7} M.Yu. Khlopov, ``PHYSICS OF DARK MATTER IN THE LIGHT OF DARK ATOMS'', {\it Modern Physical Letters A}, vol. 26, no. 38, pp. 2823-2839, 2011.

\bibitem{8} M.Yu. Khlopov, ``Introduction to the special issue on indirect dark matter searches'', {\it International Journal of Modern Physics A}, vol. 29, article 1443002, 2014.

\bibitem{9} J.R. Cudell and M. Khlopov, ``Dark atoms with nuclear shell: A status review'', {\it International Journal of Modern Physics D}, vol. 24, article 1545007, 2015.

\bibitem{9a} M. Buchkremer and A. Schmidt, ``Long-lived Heavy Quarks: A Review'', {\it Advances in High Energy Physics}, vol. 2013, Article ID 690254, 17 pages, 2013.

\bibitem{10} Yu.N. Bazhutov, G.M. Vereshkov, and V.I. Kuksa, ``Experimental and Theoretical Premises of New Stable Hadron Existence'', {\it International Journal of Modern Physsics A},
vol. 2, article 1759188, 2017.

\bibitem{16} J.C. Pati and A. Salam, ``Lepton number as the fourth colour'', {\it Physics Review D}, vol. 10, no. 1, pp. 275-289, 1974.

\bibitem{11} V. Barger, N.G. Deshpande, et al., ``Extra fermions in $E_6$ superstring models'', {\it Physical Review D}, vol. 33, no. 7, pp. 1902-1924, 1986.

\bibitem{12} V.D. Angelopoulos, J. Ellis, H. Kowalski et al., ``Search for new quarks suggested by superstring'', {\it Nuclear Physics B}, vol. 292, pp. 59-92, 1987.

\bibitem{13} P. Langacker and D. London, ``Mixing between ordinary and exotic fermions'', {\it Physical Review D}, vol. 38, no. 3, pp. 886-906, 1988.

\bibitem{14} V.A. Beylin, G.M. Vereshkov and V.I. Kuksa, ``Mixing of singlet quark with standard ones and the properties of new mesons'', {\it Physics of Atomic Nuclei}, vol. 55, no. 8,
pp. 2186-2192, 1992.

\bibitem{15} R. Rattazzi, ``Phenomenological implications of a heavy isosinglet up-type quark'', {\it Nuclear Physics B}, vol. 335, pp. 301-310, 1990.

\bibitem{10.9} M.Yu. Khlopov and R.M. Shibaev, ``Probes for 4th Generation Constituents of Dark Atoms in Higgs Boson Studies at the LHC'', {\it Advances in High Energy Physics},
article ID 406458, 7 pages, 2014.

\bibitem{13a} M. Maltoni, V.A. Novikov, L.B. Okun et al., ``Extra quark-lepton generations and precision measurements'', {\it Physics Letters B}, vol. 476, no. 1-2, pp. 107-115, 2000.

\bibitem{Ilyin} V.A. Ilyin, M. Maltoni, V.A. Novikov et al., ``On the search for 50 GeV neutrinos'', {\it Physics Letters B}, vol.  503, no. 1-2, pp. 126-132, 2001.

\bibitem{Novikov} V.A. Novikov, L.B. Okun, A.N. Rozanov and M.I. Vysotsky, ``Extra generations and discrepancies of electroweak precision data'', {\it Physics Letters B}, vol. 529, no. 1-2,
pp. 111-116, 2002.

\bibitem{Eberhardt} O. Eberhardt, G. Herbert, H. Lacker et al., ``Impact of a Higgs Boson at a Mass of 126 GeV on the Standard Model with Three and Four Fermion Generations'',
{\it Physics Review Letters}, vol. 109, no. 24, article 241802, 2012.

\bibitem{Botella} F.J. Botella, G.C. Branco, M. Nebot, ``The Hunt for New Physics in the Flavor Sector with up vector-like quarks'', {\it Journal of High Energy Physics},
vol. 12, article 040, 2012.

\bibitem{Kumar} A. Kumar Alok, S. Banerjee, D. Kumar, S.U. Sankar, D. London, ``New-physics signals of a model with vector singlet up-type quark'',
{\it Physical Review D}, vol. 92, no. 1, article 013002, 2015.

\bibitem{Llorente} J. Llorente, B. Nachman, ``Limites on new coloured fermions using precision data from Large Hadron Collider'', arXiv:1807.00894(hep-ph).

\bibitem{Peskin} M.E. Peskin, T. Takeuchi, ``Estimations of oblique electroweak corrections'', {\it Physical Review D}, vol. 46, no.1, p. 381, 1992.

\bibitem{Burgess} C. P. Burgess, S. Godfrey, H. Konig, et. al., ``A global Fit to Extended Oblique Parameters'', {\it Phys. Lett. B}, vol. 326, no. 3-4, pp. 276-281, 1994.

\bibitem{PDG} M. Tanabashi et al. (Particle Data Group), ``The Review of Particle Physics (2018)'', {\it Physical Review D}, vol. 98, 2018.

\bibitem{10.8} K. Belotsky, M. Khlopov and K. Shibaev, ``Stable quarks of the 4th family?'', arXiv:astro-ph/0806.1067 (28 pages).

\bibitem{13d} S.G. Glashow, ``A Sinister Extension of the Standard Model to $SU(3)\times SU(2)\times SU(2)\times U(1)$'', arXiv:hep-ph/0504287 (9 pages).

\bibitem{Smith} P.F. Smith, J.R.J. Bennet, G.J. Homer et al., ``A search for anomalous hydrogen in enriched $D_2O$, using a time-of-flight spectrometer'', {\it Nuclear Physics B},
vol. 206, no. 3, pp. 333-348, 1982.

\bibitem{Muller} P. Muller, L.-B. Wang, J. Holt et al., ``Search for Anomalously Heavy Isotopes of Helium in the Earth's Atmosphere'', {\it Physical Review Letters}, vol. 92, article 022501,
2004.

\bibitem{18} G.M. Vereshkov and V.I. Kuksa, ``$U(1)SU(3)$-gauge model of baryon-meson interactions'', {\it Physics of Atomic Nuclear}, vol. 54, no. 6(12), pp. 1700-1704, 1991.

\bibitem{SEeff} J.L. Feng, M. Kaplinghat, and Hai-Bo Yu, ``Sommerfeld Enhancement for Thermal Relic Dark Matter'', {\it Physical Review D}, vol. 82, article 083526, 2010.

\bibitem{Khoze} V.V. Khoze, A.D. Plascencia, K. Sakurai, ``Simplified models of dark matter with a long-lived co-annihilation partner'', arXiv: 1702.00750[hep-ph].

\bibitem{Asadi} P. Asadi, M. Baumgart, P.J. Fitzpatric et al., ``Capture and decay of EW WIMPonium'', {\it arXiv: 1610.07617 [hep-ph]}.

\bibitem{RanHuo} Ran Huo, S. Matsumoto, Y-L.S. Tsai, T.T. Yanagida, ``A scenario of heavy but visible baryonic dark matter'', {\it JHEP 1609}, p. 162, 2016.

\bibitem{Cirelli} M. Cirelli, A. Strumia, and M. Tamburini, ``Cosmology and Astrophysics of Minimal Dark Matter'', {\it IFUP-TH/2007-12}; arXiv:0706.4071[hep-ph].

\bibitem{Blum} K. Blum, R. Sato, and T.R. Slatyer, ``Self-consistent calculation of the Sommerfeld enhancement'', {\it Journal of Cosmology and Astrophysics}, vol. 06, article 021,
2016.

\end{thebibliography}
\end{document}